# A physical perspective to understand myelin. II. The physical origin of myelin development

**Authors**: Yonghong Liu[1], Yapeng Zhang[1], Wenji Yue[1], Ran Zhu[1], Tianruo Guo[3], Fenglin Liu[1], Yubin Huang[1], Tianzhun Wu*[1,2], Hao Wang*[1,2]

**Affiliations:**

[1]Institute of Biomedical & Health Engineering, Shenzhen Institutes of Advanced Technology (SIAT), Chinese Academy of Sciences (CAS), Shenzhen 518035, China

[2]Key Laboratory of Health Bioinformatics, Chinese Academy of Sciences

[3]Graduate School of Biomedical Engineering, University of New South Wales, Sydney, NSW2052, Australia

***Hao Wang** hao.wang@siat.ac.cn

***Tianzhun Wu** tz.wu@siat.ac.cn

**Abstract**

The physical principle of myelin development is obtained from our previous study by explaining Peter's quadrant mystery: an external applied negative and positive E-field can promote and inhibit the growth of the inner tongue of the myelin sheath, respectively. In this study, this principle is considered as a fundamental hypothesis, named Hypothesis-E, to explain more phenomena about myelin development systematically. Specifically, the g-ratio and the fate of the Schwann cell's differentiation are explained in terms of E-field. Moreover, an experiment is proposed to validate this theory.

## Introduction

Myelin is an insulating sheath forming around axons. Its biological function in neural systems and the growing mechanism have attracted increasing attention in the field of neuroscience [1-6]. Previous studies reported a series of experimental observations about the micro-structures of myelin. For example, 1) The spiraling directions of neighboring myelin sheaths has a certain pattern. That is, the neighboring myelin sheaths on the same axon have the same spiraling direction [7], while the neighboring myelin sheaths on the adjacent axons have the opposite spiraling directions [8-10]; 2) For oligodendrocytes (OLs), the inner and outer tongues tend to be located within the same radial quadrant [11-15]; 3) The axons of varying calibers tend to have myelin sheaths of the same thickness, resulting in the g-ratio phenomenon [16-18]; 4) Only the axon with sufficient caliber can be myelinated, resulting in the radial sorting phenomenon [19-21]; and 5) For Schwann cells (SCs), one SC can only myelinate one axon. If the SC forms the remak bundle, it can never form the myelination even if a large axon is ensheathed [21]. These experimental observations indicate a multifaceted mechanism underlying myelin growth. For example, 1) the non-random spiraling phenomenon suggests that myelin growth can be influenced by the interaction between spatially closed myelin sheaths. 2) The same quadrant phenomenon indicates that myelin growth can be influenced by the relationship between the inner and outer tongues. 3) The g-ratio phenomenon indicates a possible correlation between the inner tongue growth and the number of myelin lamellae. 4) The radial sorting phenomenon indicates a possible correlation between the myelin growth and the curvature of axons. 5) The characteristic SC properties in myelination and the remak bundle indicate the effect of the surrounding environment in formulating the growth of the inner tongue. Previous studies about mechanisms underlying myelin growth mainly focused on studying contributions from different molecules or proteins [2,19,21-26] without providing reasonable explanations for these phenomena. Thus a substantial body of new knowledge is to be discovered. In this study, we conducted in silico investigations to explore the physical origins of the reported myelin observations as follows. Our previous theoretical model [4] of Peter's quadrant mystery [11-15] reported that the growth of myelin could be guided by the electric field (E-field). We name it Hypothesis-E, "E" refers to "electric." In Hypothesis-E, an external negative E-field promotes the growth, while an external positive E-field inhibits the growth (Figure 1(a)). This study proposed three new hypotheses based on Hypothesis-E to further explain the physical origins of a series of morphological characteristics (Figure 1(b-d)) of the myelin.

## 1. Hypothesis-$E_N$ to explain g-ratio

### 1.1 g-ratio

The myelin g-ratio, defined as the ratio between the inner and the outer diameter of the myelin sheath, has been reported in many experimental studies [16-18]. This precise relationship between axonal diameter and myelin sheath thickness has been reported is one of the most enigmatic questions: how is the myelinating glial cell instructed to make precisely the correct number of wraps? Transplantation of oligodendrocytes into nerve tracts containing axons of different sizes demonstrates that the number of wraps is determined by the axon but not by the glial cell because the transplanted glial cells elaborate myelin sheaths appropriate for their new location [27]. A key axonal signal for regulating myelin sheath thickness, the growth factor neuregulin (Ngr1), is now identified by Michailov et al. [28]. However, the detailed mechanisms of controlling the myelin wrapping by the axonal signal remain unclear.

### 1.2 Hypothesis-$E_N$

The cross-section of a myelinated axon in the resting state (no action potential is activated) is shown in Figure 2(a). The intracellular potential is more negative than the extracellular potential, resulting in a negative E-field on the inner tongue. This negative E-field is the driven force making the inner tongue grow and wrap around the axon to form myelination. Then the Hypothesis-$E_N$ ("N" refers to "negative") is described as follow:

*The inner tongue of myelin is driven by a negative E-field from the axon in the resting state. The strength of the E-field on the inner tongue is proportional to its growth rate. When the E-field is lower than a threshold, the growth of the inner tongue terminates.*

## 1.3 Modeling the relationship between g-ratio and the E-field

Figure 2(b) showed a mature myelinated axon with the number of myelin lamellae as $N$. The axonal radius is $a$, and the thickness of a single myelin lamella is $b$. Then the total myelin thickness, $D$, is $b \times N$. We assume that the axonal cross-section is centrally symmetric. So we only simulate the local axon with a radial angle as $\theta$, as shown in Figure 2(b). The capacitance, $C$, of each layer is proportional to its area. Since the longitudinal length of each layer is identical, the capacitance of each layer is proportional to the arc length $l$:

$$C \propto Area \propto l$$

Then for the $n^{th}$ layer, the capacitance, $C_n$, is proportional to its arc length $l_n$:

$$C_n \propto l_n = \theta \times (a + (n-1)b)$$

The voltage, $V_n$, on the $n^{th}$ layer is $V_n = \frac{Q_n}{C_n}$.

Here $Q_n$ is the charge on the capacitor. So the voltage, $V_1$, on the first layer is $V_1 = \frac{Q_1}{C_1}$.

Since all capacitors are connected in series, as shown in Figure 2(c), the two boundary conditions are:

1. The charge on each capacitor is the same, assigned with the value of $Q$:

$$Q = Q_1 = Q_2 = Q_3 = \cdots = Q_N$$

2. The resting potential, $V_R$, is equivalent to a voltage source connected with these series-connected capacitors, as shown in Figure 2(c), so $V_R$ is the sum of the voltage on all capacitors:

$$V_R = \sum_{n=1}^{N} V_n = \sum_{n=1}^{N} \frac{Q_n}{C_n} = Q \times \sum_{n=1}^{N} \frac{1}{C_n}$$

The charge, $Q$, on each capacitor is:

$$Q = \frac{V_R}{\sum_{n=1}^{N} \frac{1}{C_n}}$$

The voltage, $V_1$, on the first layer, which is the inner tongue, is as shown below:

$$V_1 = \frac{Q}{C_1} = \frac{V_R}{C_1 \times \sum_{n=1}^{N} \frac{1}{C_n}} = \frac{V_R}{a \times \sum_{n=1}^{N} \frac{1}{(a+(n-1)b)}} \quad (1)$$

when the voltage potential, $V_R$, and the thickness of a single myelin lamella, $b$ are constants, the voltage on the inner tongue, $V_1$, is only a function of the number of layers $N$, axonal radius $a$, and monotonically decreases with the number of layers, $N$. Here the threshold E-field proposed in Hypothesis-$E_N$ is defined as $V_{N-T}$ ("N" refers to "negative" and "T" refers to "threshold"). And the ratio between $V_{N-T}$ and $V_R$ is defined as $\eta_{N-T}$:

$$\eta_{N-T} = \frac{V_{N-T}}{V_R} \quad (2)$$

Then the criteria for the max number of myelin lamellae $N_{max}$ is:

$$\begin{cases} V_1 \geq V_{N-T} & when\ N = N_{max} \\ V_1 \leq V_{N-T} & when\ N = N_{max} + 1 \end{cases} \quad (3)$$

Substitute (1) and (2) into (3) and get:

$$\begin{cases} \dfrac{1}{a \times \sum_{n=1}^{N_{max}} \dfrac{1}{(a+(n-1)b)}} \geq \eta_{N-T} \\ \dfrac{1}{a \times \sum_{n=1}^{N_{max}+1} \dfrac{1}{(a+(n-1)b)}} \leq \eta_{N-T} \end{cases} \quad (4)$$

(4) can be further simplified as follow:

$$\dfrac{1}{a \times \sum_{n=1}^{N_{max}} \dfrac{1}{(a+(n-1)b)}} \approx \eta_{N-T} \quad (5)$$

As seen, the value $N_{max}$ is a function of $a$ and $\eta_{N-T}$, while $b$ is constant:

$$N_{max} = f_1(a, \eta_{N-T})$$

Then *g-ratio* is also a function of $a$ and $\eta_{N-T}$:

$$g_{ratio} = \dfrac{a}{a+D} = \dfrac{a}{a+b \times N_{max}} = \dfrac{a}{a+b \times f_1(a,\eta_{N-T})} = f_2(a,\eta_{N-T})$$

To enable calculating these two functions, we need to obtain the constant of $b$. Based on previous studies, we set $b$=17 nm as a typical value [29]. The g-ratio and $N_{max}$ simulation is shown in Figures 3(a) and (b).

### 1.4 Results

In Figure 3(a), the g-ratio curve monotonically increases with axonal radius $a$. The curve of $N_{max}$ has a decreasing slope with $a$, approaching a constant value determined by $\eta_{N-T}$. It is emphasized that $N_{max}$ is the maximum number of myelin lamellae of an axon. The actual measured number of myelin lamellae $N$ should be no more than $N_{max}$ shown in Figure 3(b), $N \leq N_{max}$. Thus, g-ratio curves shown in Figure 3(a) is a minimum value, which is a lower limit. As illustrated in Figure 3(d), all measured data points of the g-ratio shall be higher than the g-ratio curve in Figure 3(a). By fitting the curve of this lower edge, the actual $\eta_{N-T}$ can be obtained. In Figure 3(e), we validate our simulations with the experimental data from previously published studies of myelinated axons [30-33,18,34-43,60]. We found a clear edge can be formed (the red fitting curves are plotted by ourselves for an indicator of boundary). As mentioned above, by fitting this lower edge, the threshold voltage, which is an important characteristic of the target nervous system, can be obtained. This characteristic is not recognized yet in conventional theories and models. Noticeably, $N_{max}$ goes infinite when $a$ approaches zero, indicating that the axon with a very small diameter can have infinitely thick myelin. However, the axons with a radius within the divergence region in Figure 3(b) are unmyelinated. We will make a more detailed discussion in the next section.

### 1.5 Discussion

#### 1.5.1 Why does the divergence happen?

The condition to achieve $N_{max}$ is to meet the condition of equation (5), as written again here:

$$\dfrac{1}{a \times \sum_{n=1}^{N_{max}} \dfrac{1}{(a+(n-1)b)}} \approx \eta_{N-T} \quad (5)$$

However, the limit of $\eta_{N-T}$ when $N_{max}$ approaches infinite is as follow:

$$\lim_{N_{max} \to \infty} \eta_{N-T} = \lim_{N_{max} \to \infty} \dfrac{1}{a \times \sum_{n=1}^{N_{max}} \dfrac{1}{(a+(n-1)b)}} = \dfrac{b}{a}$$

where $b/a$ is the lower limit of $\eta_{N-T}$. If the actual $\eta_{N-T}$ is above this lower limit, $V_1$ is lower than $V_{N-T}$ when $N_{max}$ is a finite number; then the myelin growth stops (eq (6)). However, if the actual is $\eta_{N-T}$ lower than this lower limit, $V_1$ can never be reduced to $V_{N-T}$, whatever $N_{max}$ is; the myelin growth never stop (eq(7)).

$$\text{when } \frac{b}{a} < \eta_{N-T}; N = finite\ number \quad (6)$$

$$\text{when } \frac{b}{a} \geq \eta_{N-T}; N \to \infty \quad (7)$$

where the occurrence of divergence is determined by the ratio between the thickness of single-layer myelin, *b*, and the axonal radius, *a*. When *a* is large enough to meet eq (6), the calculation of $N_{max}$ is convergent. Otherwise, when *a* is a small number, which is the case of unmyelinated axons, the divergence happens. A more intuitive modeling result is shown in Figure 3(c). Since $V_I$ decreases with the growth of myelin lamellae, the ratio of $V_I/V_R$ will decrease with *N*. Then this ratio reaches the value of $\eta_{N-T}$, the curve stops at the value of $N_{max}$. As seen, the curve of the axonal diameter of 0.8 μm, 1.4 μm, 2.6 μm and 6.2 μm can have a finite value of $N_{max}$. However, when axon diameter is 0.2 μm, the curve of $V_I/V_R$ approaches $\frac{b}{a} = 0.17$ (*b*=17nm and axonal radius *a*=100nm), which is higher than $\eta_{N-T}$ =1/18≈ 0.056, the growth cannot be stopped.

### 1.5.2 The relation between the divergence region and unmyelinated axons

Since the modeling result can closely predict the biological observations of the g-ratio and myelin thickness at different axonal diameters, we tend to explore the biological meaning hidden behind the divergence region. It is observed that the number of myelin lamellae suddenly decreases to zero when the axonal diameter is lower than a threshold, indicating that some unknown factors dominant the growth of smaller myelin and forbid the process of myelination.

Interestingly, Hypothesis-$E_N$ suggests that the axon of a very small diameter can have infinitely thick myelin, which disagrees with biology. Therefore, during myelin development, some unknown factor that inhibits myelin growth is introduced when the axonal diameter is lower than a certain value. We will discuss this unknown factor in the section Hypothesis-$E_D$.

### 1.5.3 An introspection of this model

The origin of the g-ratio is the myelin's growth rate inversely proportional to its layers. That is, the promoting factor of myelin growth decays with its layers. Meanwhile, the inner tongue is the growing terminal of the myelin, indicating this promoting factor exerts its function on the inner tongue. In our model, the voltage, $V_I$, on the inner tongue meets these boundary conditions. Any alternative theories shall also meet the above-mentioned boundary conditions. Since this $V_I$ is obtained from Hypothesis-$E_N$, so it is renamed as $V_{EN}$ in this article to avoid confusion.

## 2. Hypothesis-$E_D$ to explain radial sorting

### 2.1 Radial sorting

Radial sorting is the process by which Schwann cells choose larger axons to myelinate during development [61]. During this process, SCs proliferate and expand cellular extensions into bundles of unsorted axons to detach individual axons and establish the 1:1 relationship (one SC can only myelinated one axon) required for myelination [44]. Axons with a diameter of <1 μm remain in bundles, and SCs in contact with these axons differentiate into unmyelinated SCs, called remak bundles [45]. This radial sorting process is reported to be tightly regulated and depends on signals from axons as well as the extracellular matrix [46].

### 2.2 Hypothesis-$E_D$

With the radial sorting process, SCs can recognize functionally-identified axons just by their calibers. Axons of large caliber possess a promoting factor, while the axons of small caliber possess an inhibiting factor to the myelin growth. Based on Hypothesis-E, we can predict that larger axons can possess a more negative voltage than that of smaller axons. Then Hypothesis-$E_D$ ("D" refers to "dipole") refers to:

*When SCs get close to the surfaces of axons, axons of larger caliber will exert a special "E-field," which is more negative than that of the axons of smaller caliber, to the cell membrane of SCs. Thus, SCs tend to grow and wrap on larger axons. When the caliber of axons is lower than a threshold, the amplitude of the negative E-filed is too low to enable the growth of SCs on their surfaces.*

## 2.3 Modeling the relationships between the radial sorting and the dipole potential

In Figure 4(a), the axon membrane lipid bilayer consists of two layers of amphiphilic molecules. The positively charged hydrophobic tails of these lipids are directed toward the membrane center, while the negatively charged hydrophilic heads are directed toward the extra- and intracellular fluid [63]. Each amphiphilic molecule is an electric dipole, a group of separated charges with opposite polarities. The potential, also called the dipole potential, generated by this lipid bilayer is shown in Figure 4(a). Such dipole potential has two negative peaks at the extra- and intracellular surface and one positive peak at the membrane center [47-50]. The bending of the cell membrane will break the centrosymmetry of the bilayer structure and change the amplitude of those two negative peaks, which is called the flexoelectric effect [51]. In particular, the amplitude of the left negative peak located at the extracellular surface, named $V_{D-N1}$ ("D" refers to "dipole" and "N" refers to "negative"), decreases with bending, while the amplitude of the right negative peak located at the intracellular surface, named $V_{D-N2}$, increases with bending. When the SC membrane contacts with the axon surface, a portion of $V_{D-N1}$, labeled as $\Delta V_{D-N1}$ in Figure 4(b), is applied across SC's membrane. This $\Delta V_{D-N1}$ meets the criteria of growth promotion, which is an external negative E-field. Meanwhile, the amplitude of this $\Delta V_{D-N1}$ increases with the axon caliber and saturates at a certain value (Figure 4(c)). The detailed modeling and calculation process of the dipole potential can be found in Supplementary. Interestingly, $\Delta V_{D-N1}$ has a sudden decline from the position of about 400 nm, which is roughly the threshold of radial sorting of SCs.

The surface potential of the cell membrane can influence the binding affinity of the peptide to lipid bilayers [52]. So it is conjectured that the binding affinity between the polarized protein molecules on the SC membrane and axons, which are responsible for the interface adhesion, is positively correlated with the surface dipole potential of the axon $\Delta V_{D-N1}$. When the axon caliber is large, $\Delta V_{D-N1}$ is strong enough for the molecules to form the bound; thus, SCs can grow and wrap on these axons to form myelin. However, when the axon caliber is lower than a certain value, e.g., 400 nm in Figure 4(c), $\Delta V_{D-N1}$ is insufficient to provide the binding affinity, leading to the failure of SC in adhering to the axon. These simulation results suggest that the dipole potential from the axon surface can be one of the factors influencing myeline developments.

## 2.4 An introspection of this model

In radial sorting, SCs can robustly identify the biologically-identical axons merely by their physical calibers. Meanwhile, during the contact with the axon of difference calibers, the only difference can be experienced by SCs is the curvature. So it can be inferred that this physical identification signal is related to the surface curvature. To the best of our knowledge, the dipole potential is the sole physical factor determined by the axon caliber and whose changing trend is consistent with Hypothesis-E. Since this $\Delta V_{D-N1}$ is obtained from Hypothesis-$E_D$, it is renamed as $V_{ED}$ to avoid confusion with other variables.

## 3. Hypothesis-$E_P$ to explain behaviors of SCs

### 3.1 Different behaviors of SCs in myelination and remak bundle

The SCs behave differently in myelination and remak bundles [19]. In the scenario of myelination, an SC will wrap around a large axon with a 1:1 relationship. In the scenario of a remak bundle, an SC can never form myelination, even if a large axon is ensheathed.

### 3.2 Hypothesis-$E_P$

Hypothesis-$E_P$ (P refers to "Positive") is proposed to reveal the myelination criteria and explain the mechanism underlying differential SC activities:

*The growth of the inner tongue of myelin is inhibited by a positive E-field induced by action potentials. The strength of the E-field on the inner tongue is proportional to its capability of growth-inhibiting. When the E-field is lower than a threshold, it does not exert its inhibition function.*

In Figure 5(a), a new perspective about how the myelin growth is modulated by E-field is shown. $V_P$ and $V_N$ refer to the amplitude of resting potential and the positive peak voltage of the action potential, respectively. The threshold voltage $V_{P-T}$ is the threshold voltage to inhibit myelin growth, while $V_{N-T}$ is the threshold voltage to promote myelin growth. The ratio between $V_{P-T}$ and $V_P$ is $\eta_{P-T}$, and the ratio between $V_{N-T}$ and $V_N$ is $\eta_{N-T}$. The area higher than $V_{P-T}$ is the inhibition phase (red area in Figure 5(a)), while the area lower than $V_{P-T}$ is the promotion phase of myelin growth (blue area in Figure 5(a)).

In Figure 5(b-i, ii), the total voltage $V$ (this voltage can be either the resting potential $V_R$ or the action potential $V_A$, "$A$" refers to "action") across a single-layer myelin is applied on $C_{1-A}$ and $C_{2-A}$ ("A" refers to the capacitor of case A in Figure 5(b)):

$$C_{1-A} \propto a;$$

$$C_{2-A} \propto 2 \times (a + b);$$

$$Q_{1-A} = C_{1-A} \times V_{1-A} = Q_{2-A} = C_{2-A} \times V_{2-A} = Q;$$

$$V_{1-A} + V_{2-A} = V;$$

Since $C_{2-A}$ only has a single layer of the cell membrane, the equivalent capacitance shall be doubled compared with the one with double layers of the cell membrane.

Then the ratio between the voltage on $C_{1-A}$ and $V$ is:

$$\eta_A = \frac{V_{1-A}}{V} = \frac{1}{1+\frac{a}{2a+2b}} = \frac{1}{1+\frac{1}{2+2\times\frac{b}{a}}};$$

In Figure 5(b-iii, iv), the total voltage $V$ across a double-layer myelin is applied on $C_{1-B}$, $C_{2-B}$, and $C_{3-B}$ ("B" refers to the capacitor of case B in Figure 5(b)):

$$C_{1-B} \propto a;$$

$$C_{2-B} \propto a + b;$$

$$C_{3-B} \propto 2 \times (a + b + b');$$

$$Q_{1-B} = C_{1-B} \times V_{1-B} = Q_{2-B} = C_{2-B} \times V_{2-B} = Q_{3-B} = C_{3-B} \times V_{3-B} = Q;$$

$$V_1 + V_2 + V_3 = V;$$

Here we set the thickness of the second layer is $b'$, which is different from that of the first layer $b$. Since $C_{3-B}$ only has a single layer of the cell membrane, its equivalent capacitance shall be doubled.

Then the ratio between the voltage on $C_{1-B}$ and $V$ is:

$$\eta_B = \frac{V_{1-B}}{V} = \frac{1}{1+\frac{a}{a+b}+\frac{a}{2(a+b+b')}};$$

Since the myelin lamellae are not compact yet at the initial myelination process, $b$ is a value comparable with $a$. So here we set the ratio of $b/a$ is 0.1, which is a typical value and a reasonable approximation, to further simplify the equation of $\eta_A$ and $\eta_B$ as below:

$$\eta_A = \frac{1}{1+\frac{1}{2+2\times\frac{b}{a}}} \approx 0.88;$$

$$\eta_B = \frac{1}{1+\frac{a}{a+b}+\frac{a}{2(a+b+b')}} = \frac{1}{1.909+\frac{1}{2(1.1+\frac{b'}{a})}};$$

As seen, $\eta_A$ is a constant, meaning that about 88% of the transmembrane voltage, which can be either $V_R$ or $V_A$, will be applied onto the adaxonal layer of the myelin. Meanwhile, $\eta_B$ is a function of $\frac{b'}{a}$, which is calculated as shown in Figure 6(c). $\eta_B$ increases with $\frac{b'}{a}$.

Then let's consider the situations of the wrapping of the second myelin lamella on a large axon by both a normal SC and a remak SC, as shown in Figure 5(b) and Figure 6(a), respectively. For an SC forming myelination, the condition is similar to Figure 5(b) when $a/b'$. So its $\eta_B$ is located within the blue region in Figure 6(c), labeled with myelination region. For a remak SC, the condition is similar to Figure 6(a). When a large axon is ensheathed by a remak bundle, initially the axon is wrapped by a SC as shown in Figure 6(a-i&ii). When one of the SC terminal tends to further grow and wrap the large axon to form myelin, it inevitably faces the situation shown in Figure 6(a-iii&iv) when $b'$ is comparative or even larger than $a$. Thus its $\eta_B$ is located within a pink region in Figure 6(c), labeled with the non-myelination region.

In Hypothesis-$E_P$, a positive voltage $V_P$ in the action potential can inhibit myelin growth. Therefore, for a normal SC myelinating a large axon, the inhibiting voltage exerted upon the inner tongue is lower. The promoting factor induced by the negative voltage (mainly comes from the resting potential) dominates the myelin growth (see Figure 6(b)). However, for a remak SC, the inhibiting voltage upon the inner tongue is higher. Thus the inhibiting factor dominates the myelin growth, stopping the wrapping of the second layer. Since this inhibitory voltage on the inner tongue, $V_P \times \eta_B$, is obtained from Hypothesis-$E_P$, it is renamed as $V_{EP}$ to avoid confusion with other variables. This simulation under Hypothesis-$E_P$ support the experimental observation of the radial sorting [61], that is, 1) An SC can merely myelinate one axon. 2) Remak SC cannot myelinate. Moreover, we can also make a rough estimation of $\eta_{P-T}$. It is a value located close to the myelination and non-myelination region interface 0.43~0.46 shown in Figure 6(c).

### 3.3 Discussion

This model also indicates potential explanations for other experimental observations, as discussed below.Firstly, it is contradictory to the conventional understanding of the correlation between neural activities and myelin development. It was widely believed that the action potential is a positive factor in the myelination process [63], while in our model, it is a negative factor. If our model is correct, it can be predicted that by eliminating the action potential during myelin development, the myelin can grow thicker. Chan et al. have confirmed this hypermyelination phenomenon of oligodendrocytes by muting the action potential [53], which is supporting evidence of our model. It can be foreseen that the same phenomenon can be observed in the experiment of SCs. Secondly, the frequency of the action potential is also a factor affecting the fate of myelination. When the action potential is activated more frequently, which is the case of sensory fibers, the inhibiting factor tends to dominate, and the axons tend to be unmyelinated. Conversely, when the action potential is activated more rarely, which is the case of motor fibers, the promoting factor tends to dominate, and the axons tend to be myelinated. This may partially explain that the majority of the sensory fibers are unmyelinated while the counterparts of the motor fibers are myelinated [54]. This model also indicates a positive correlation between neural hyperactivity and the degeneration of myelin. It may provide a clue for the neurodegenerative disorders such as Parkinson's disease, whose early-stage symptoms, such as hand tremor and muscle stiffness, are the results of uncontrollable hyper-activation of some neurons, while the accompanying symptoms include the demyelination of neurons. At least, these phenomena are not contradictory to our model.

### 3.4 A recap of the g-ratio phenomenon

The observation of the hypermyelination of oligodendrocytes by muting the action potential [53] reveals the relationship between myelination and neural activities. This observation indicates that an axon with fewer action potentials tends to have thicker myelin, while an axon with more action potentials tends to have thinner myelin. So the action potential is an inhibitory factor to the myelin growth, which agrees with our theory. Meanwhile, it also indicates a quantitative relationship between the frequency of action potential and the myelin thickness. An illustrative drawing of this quantitative relationship is shown in Figure 7(a). For an axon with a very high frequency of action potential (Figure 7(a-1)), it tends to be unmyelinated, which refers to the unmyelinated region (Region 1 in Figure 7(a-1)). For an axon with a very low frequency of action potential, its myelin can grow to the maximum thickness, which refers to the lower edge (Region 3 in Figure 7(a-2)). For the axon with a medium frequency of action potential, its myelin cannot grow to the maximum thickness, even at its mature state. This scenario refers to Region 2 in Figure 7(a-2). Then based on our theory, a complete explanation of the g-ratio phenomenon, including the non-myelination region, the lower edge, and the scattering data distribution, is proposed in Figure 7(a). Moreover, our theory also makes another very interesting prediction. If unmyelinated axons tend to have a lower diameter and higher frequency of action potentials, it indicates a relationship between the action potential and the axonal diameter. In other words, the action potential is also an inhibitory factor to the radial growth of the axon.

## 3.5 An introspection of this model

Currently, we cannot claim this is the exclusively correct model. But it is highly consistent with the whole theory. The behavior of SCs is determined by whether the inner tongue can further grow to form the second layer, which is still a growth issue. Since the inner tongue growth is affect by E-field in this theory, by leveraging the same circuit model and Hypothesis-E, we can easily acquire the explanatory model shown in Figure 6 without adding any new hypotheses.

## 4. A rethinking of the complete model

At the beginning of this study, we have proposed the Hypothesis-E, which conjectures that the development of myelin is guided by an E-field applied upon the inner tongue. By explaining different phenomena of myelin development, it is concluded that this E-field consists of three components, as summarized below.

1. The component, $V_{EN}$, from $V_R$. Although $V_R$ is almost an identical value for axons of different calibers, its component, $V_{EN}$, applied to the inner tongue changes with both the axon calibre and the number of myelin lamellae, explained in Figure 2. Therefore, $V_{EN}$ is a function of both the axon caliber, $a$, and the number of myelin lamellae, $N$:

$$V_{EN} = f_{EN}(a, N);$$

2. The component, $V_{EP}$ from $V_A$. This $V_{EP}$ functions the same as $V_{EN}$ in the circuit, just with a different waveform. So it is also a function of $a$ and $N$ and changes with the same trend as $V_{EN}$:

$$V_{EP} = f_{EP}(a, N);$$

3. The component, $V_{ED}$, is from the dipole potential of the cell membrane. This component does not change with the number of myelin lamellae, $N$. So it is just a function of $a$:

$$V_{ED} = f_{ED}(a);$$

The voltage upon the inner tongue, $V_I$, is the sum of these three components:

$$V_I = f_{EN}(a, N) + f_{EP}(a, N) + f_{ED}(a); \quad (8)$$

The detailed waveform is shown in Figure 7(b).

Since the major target of this study is to establish a new theoretical framework for the mechanism of myelin development, we do not intend to involve an accurately quantitative comparison of the importance of each component. However, a very rough and qualitative analysis can still help us have a better understanding. The amplitude of the dipole potential of the lipid membrane, whose measurement is not an easy task, is estimated within the range of 200~1000 mV [55-57]. It means $V_{ED}$, which is just a small portion of the dipole potential as shown in Figure 4, may possess an amplitude of tens of mV, which is a comparative value to $V_R$ and $V_A$. Meanwhile, $V_{EN}$ and $V_{EP}$ take a small ratio of $V_R$ and $V_A$, respectively. Thus, $V_{ED}$ may take the major portion of $V_I$. In this scenario, $V_I$ has no substantial positive part. So a complete Hypothesis-E, which is a corrected version of Hypothesis-E$_P$ in Figure 5(a), is illustrated in Figure 7(b) and described below:

*The growth of the myelin is promoted by the negative E-field when it exceeds a threshold, represented by the potential $V_{N1-T}$, and inhibited by the negative E-field when it is lowered than another threshold, represented by the potential $V_{N2-T}$, respectively.*

Meanwhile, the conclusion about the g-ratio explanation in Figure 2 should also be corrected from two perspectives. Firstly, actual myelin growth is modulated by $V_I$, the sum of three component, rather than just one component assumed in Figure 2. Meanwhile, $V_{EN}$ and $V_{ED}$ has their own changing trends with the axonal diameter, it is unclear how $V_I$ changes with axonal diameter. Therefore, the actual lower limit curve may deviate from the caculated one in Figure 3. The second correction comes from the difference observations of myelin thickness. Some studies reported that the axon caliber is weakly correlated with the myelin thickness [16,17,18,36]. The number of myelin lamellae is normally lower than 50. However, we also notice that in some studies, it is reported that the myelin can have a perpetual growth, which makes the number of myelin lamellae more than 100 [58-59]. Meanwhile, in this scenario, a larger axon tends to have thicker myelin. If seems the divergence in Figure 3(b) can happen in some condition. As explained in Figure 3(c), it is because the voltage on the inner tongue ($V_{ED}+V_{EN}$) is always higher than the threshold voltage. Considering that $V_{ED}$ does not decay with the increasing number of myelin lamellae, it is highly possible that $V_{ED}$ can solely provide the voltage to promote the myelin growth, the myelin can grow perpetually with a constant growth rate, which agrees with the description in a previous study [58], quotes here:

*It is, moreover, concluded that myelin production on the average seems to be a perpetual process which, in the fully mature cat, operates at the same rate regardless of axon size.*

A possible experiment for the validation of this theory is proposed in **Supplementary S2**, in which the myelination on non-axon fibers is controlled by the applied E-field.

**Conclusion**

Our simulation suggests that myelin development can be modulated by E-field. This E-field is induced by three origins: the resting potential, the action potential, and the dipole potential. Each has its unique changing patterns with the axonal caliber and the number of myelin lamellae. Our model can be used to explain a series of observed phenomena during myelin development, such as radial sorting and g-ratio. Furthermore, our model reveals that the myelination process can be controlled by physical factors, bridging neural electrical activities and neural development.

**Acknowledgments**

This work was supported by the grant from Guangdong Research Program (2019A1515110843), Shenzhen Research Program (JCYJ20170818152810899, GJHZ20200731095206018), Chinese Academy of Sciences Research Program (2011DP173015,172644KYSB20190077) and National Natural Science Foundation of China grants(31900684).

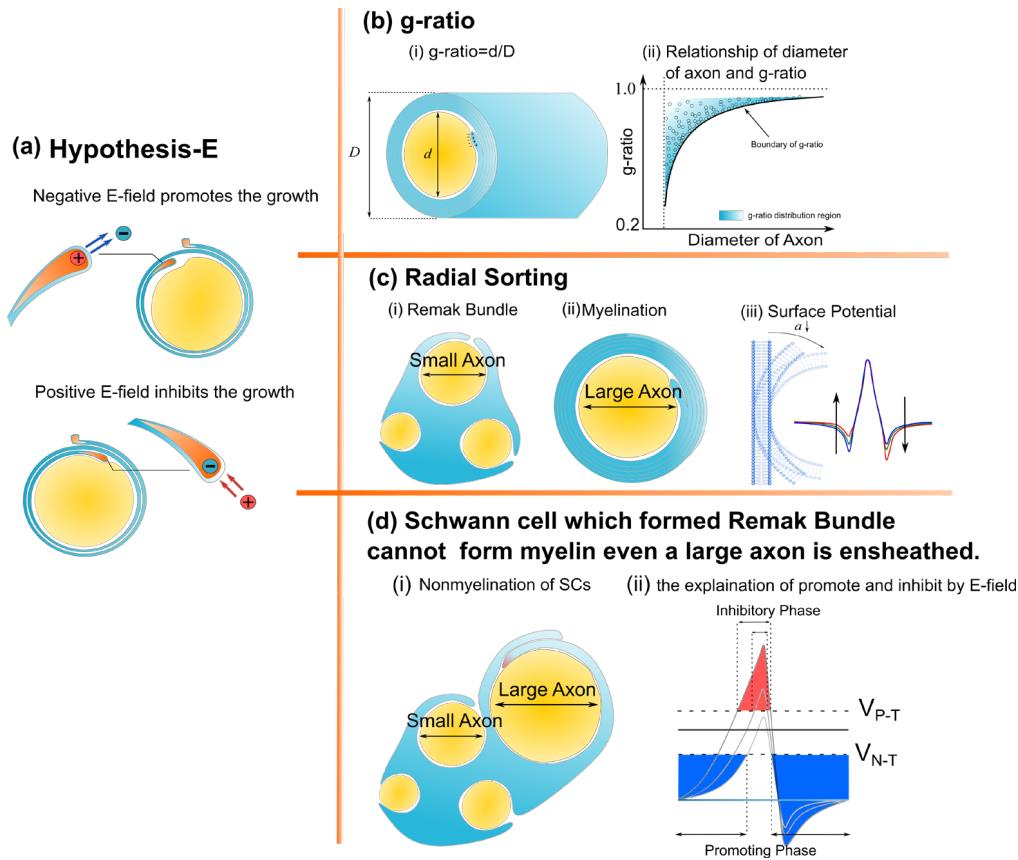

Figure 1. A series of morphological characteristics of myelin explained a mathematical and physical perspective. (a) Hypothesis-E: The effect of E-field on myelin growth; (b-c) phenomena explained by Hypothesis-E: (b) g-ratio: The thickness of myelin sheath has a specific relationship with the diameter of the axon. (c) Radial Sorting: Myelin selectively myelinated axon based on axonal diameter; (d) SC of remak bundle cannot form myelin even when a large axon is ensheathed.

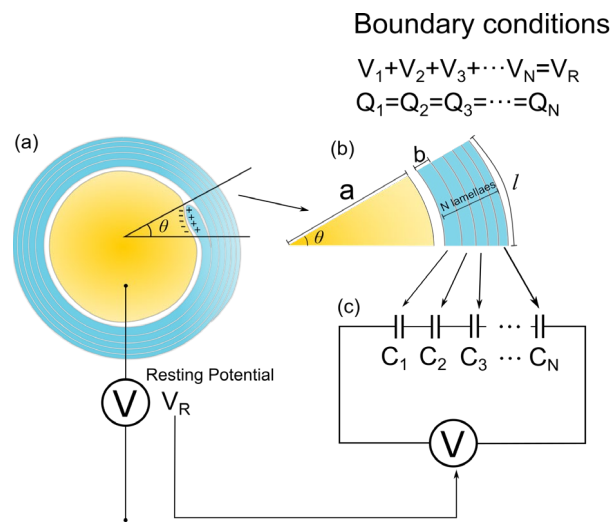

Figure 2 The model to explain g-ratio. (a) The cross-section of a myelinated axon in the static condition, the resting potential is equivalent to a voltage source; (b) A section of myelin cross-section with a radial angle of $\theta$; (c) The equivalent circuit modeling the myelin cross-section.

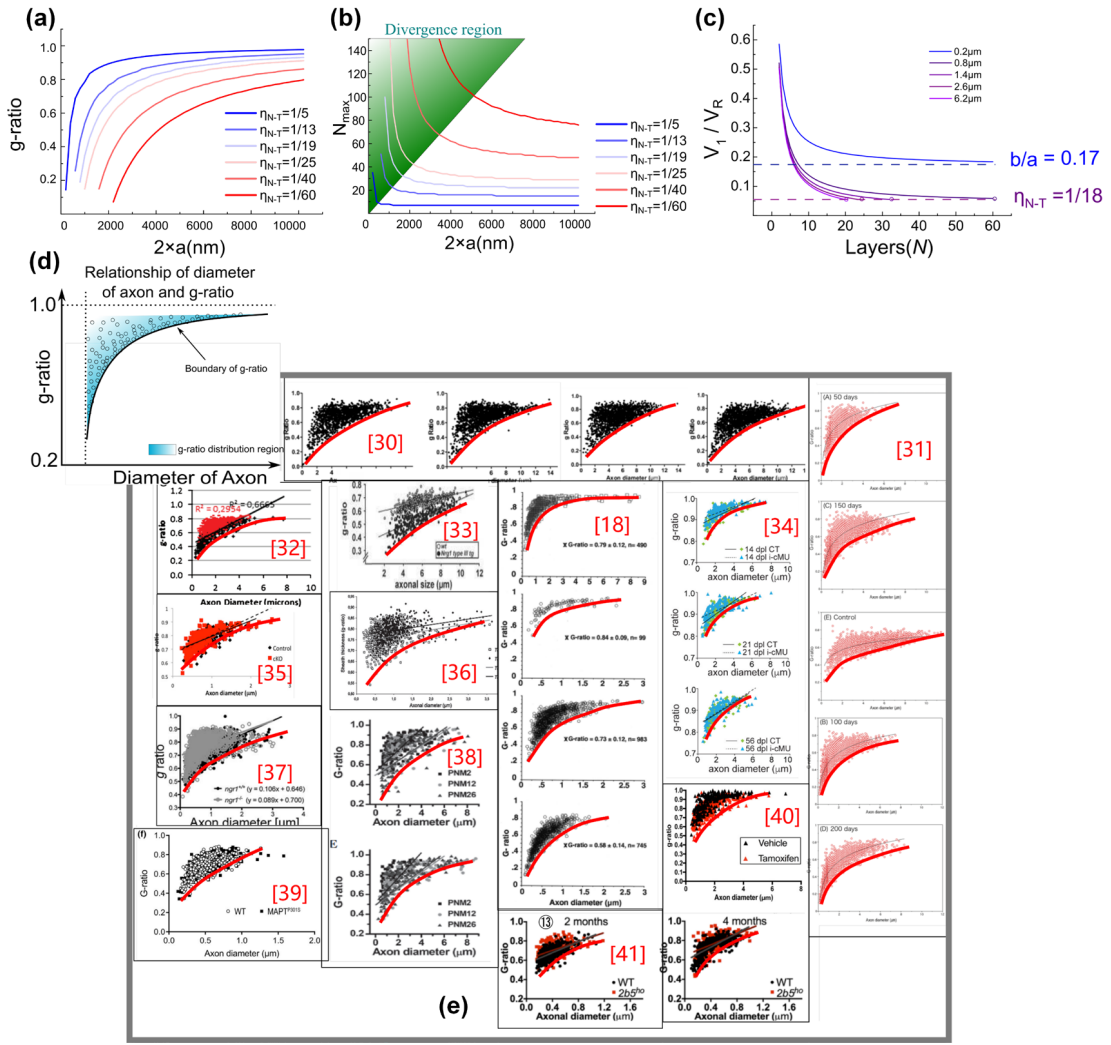

Figure 3. (a-b) g-ratio values and the maximum number of myelin lamellae, $N_{max}$ values, given different $\eta_{N-T}$. (c) The relationship between $N_{max}$ and $V_1/V_R$. When b/a<$\eta_{N-T}$, $N_{max}$ is a finite number, otherwise $N_{max}$ is infinite. (d) Illustration of the claim: the measured statistical data of g-ratio shall locate above the g-ratio curve; (e) The measured statistical data of g-ratio in publications [30-33,18,34-41].

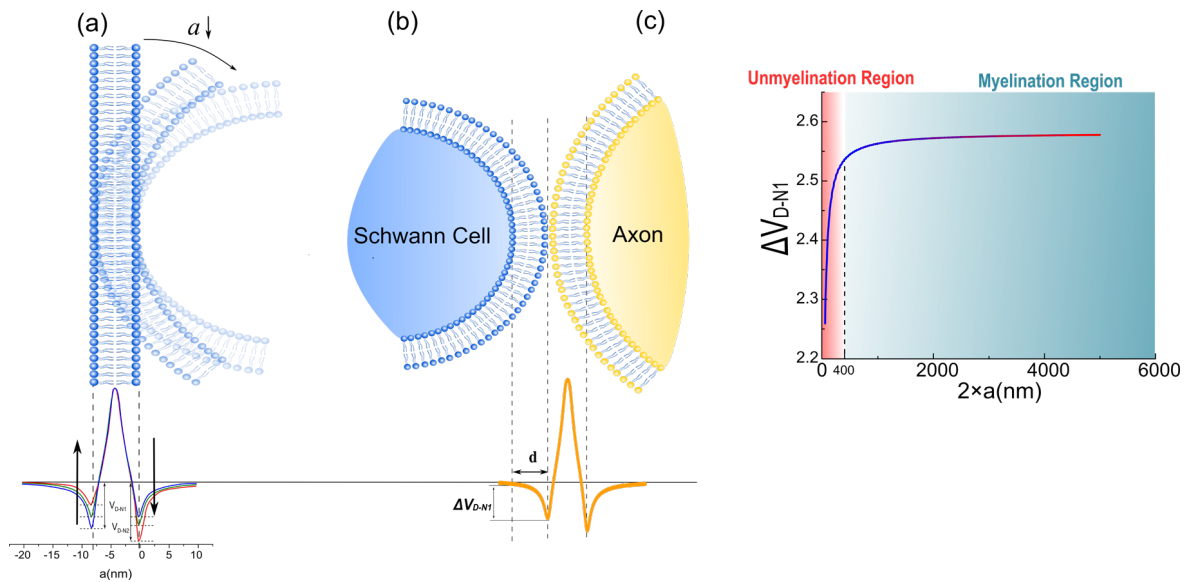

Figure 4. (a) The dipole potential generated by the bending of the cell membrane; (b) When a Schwann cell contact with the cell membrane of an axon, a portion of the surface potential, $V_{D-NI}$, will be exerted upon Schwann cell's membrane, labeled as $\Delta V_{D-NI}$; (c) The simulation result of $\Delta V_{D-NI}$ bas a function of axonal diameters.

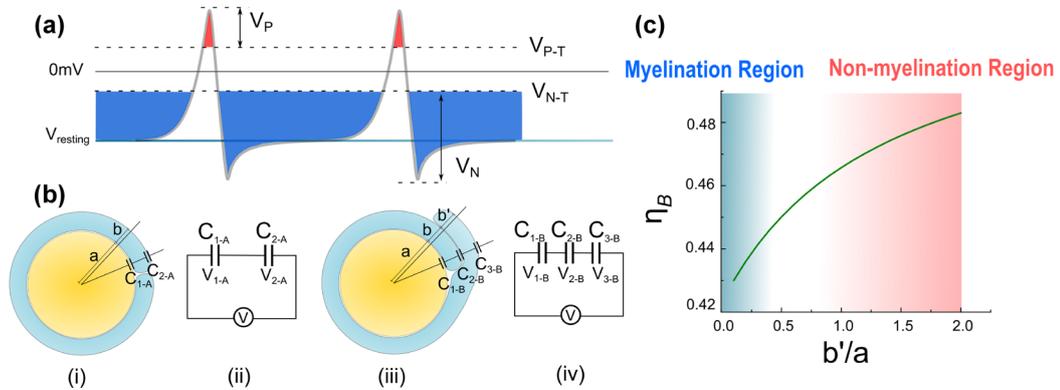

Figure 5. (a) The illustration of Hypothesis combining $E_N$ and $E_P$; b-(i),(ii) the condition of case A with one layer of myelin and its equivalent circuit. (iii),(iv) the condition of case B with double- layer myelin and its equivalent circuit. (c) Calculate result of $\eta_B$ curve changes with $\frac{b\prime}{a}$

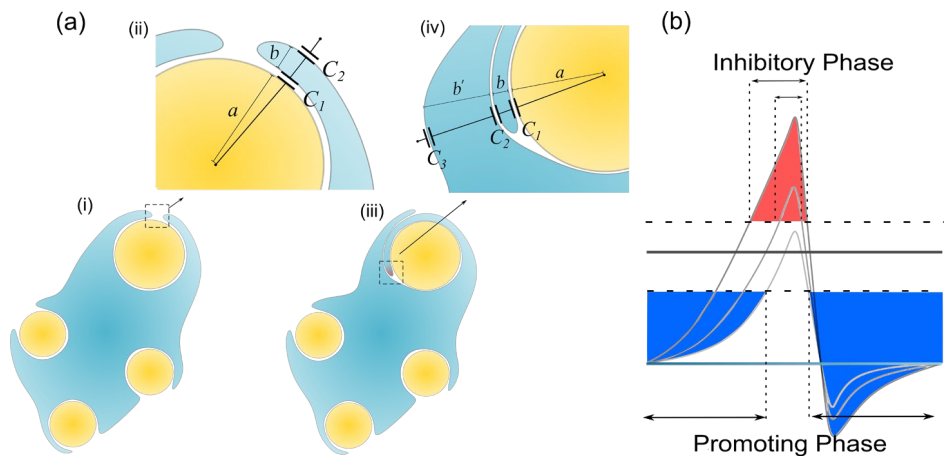

Figure 6. (a) The scenario when an SC of remak bundle ensheathes a large axon: (i-ii) only one layer of SC is wrapped; (iii-iv) when the SC tries to wrap the second layer; (b) A more decayed action potential induces a shorter inhibitory phase (red region).

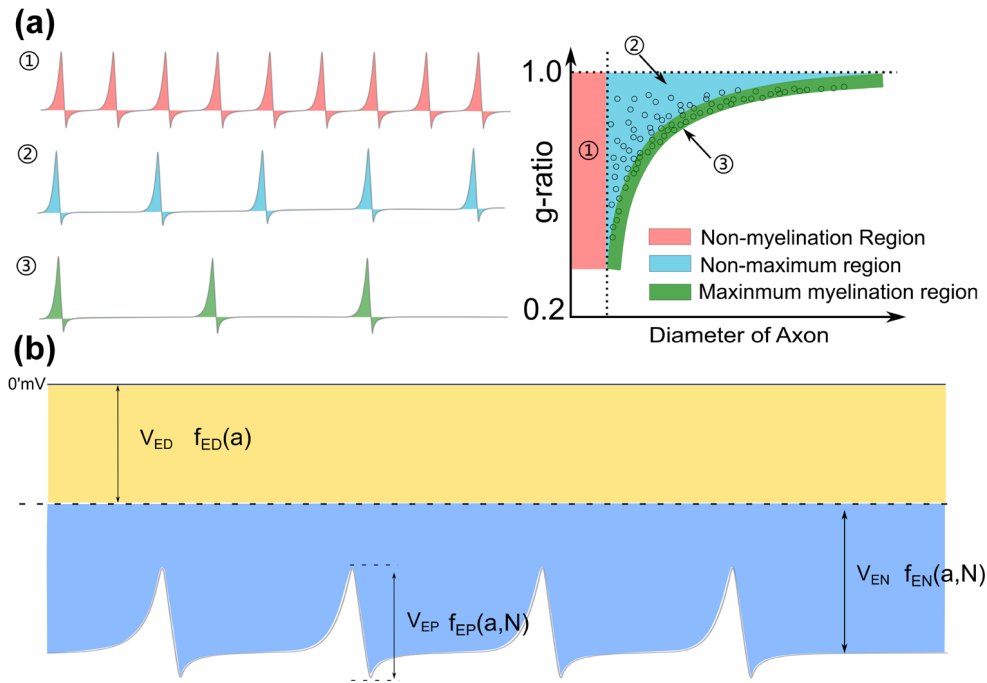

Figure 7. (a) The relationship between the frequency of action potential and the thickness of myelin: 1. An axon with a very high frequency of action potentials tends to be unmyelinated; 2. An axon with a medium frequency of action potentials cannot have maximum myelin thickness; 3. An axon with a very low frequency of action potentials tends to have maximum myelin thickness, forming the lower edge of the g-ratio data. (b) A complete perspective of Hypothesis-E: the total voltage consists of three major components: $V_{ED}$, $V_{EN}$ and $V_{EP}$.

Supplementary of

# A physical perspective to understand myelin. II. The physical origin of myelin development

**Authors**: Yonghong Liu[1], Yapeng Zhang[1], Wenji Yue[1], Ran Zhu[1], Tianruo Guo[3], Fenglin Liu[1], Yubin Huang[1], Tianzhun Wu*[1,2], Hao Wang*[1,2]

**Affiliations:**

[1]Institute of Biomedical & Health Engineering, Shenzhen Institutes of Advanced Technology (SIAT), Chinese Academy of Sciences (CAS), Shenzhen 518035, China

[2]Key Laboratory of Health Bioinformatics, Chinese Academy of Sciences

[3]Graduate School of Biomedical Engineering, University of New South Wales, Sydney, NSW2052, Australia

***Hao Wang** hao.wang@siat.ac.cn

***Tianzhun Wu** tz.wu@siat.ac.cn

## S1. The modeling process of the dipole potential of the lipid bilayer membrane

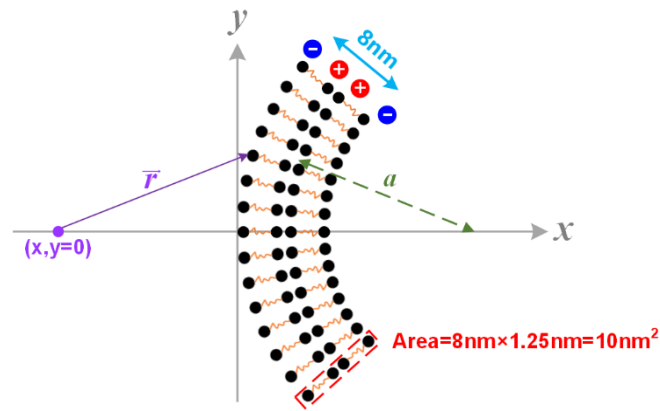

Figure S1. The modeling details of the calculation of the cell membrane's dipole potential

The detailed modeling process is illustrated in Figure S1. The arrangement of each polar is determined by the thickness of the bilayer and the diameter of the axon. Here axon radius, $a$, is a variable. The total thickness of the lipid bilayer is 8 nm, a typical value of cell membrane. The length of the dipole of each amphiphilic molecule is 3.6 nm, while the distance between the two positively charged polar is 0.8 nm. The cross-sectional area of each group of molecules is 10 nm². The charge quantity of the $n_{th}$ polar is $q_n$. The route from $n_{th}$ polar to the specific point (x,y=0) on the x-axis is $\vec{r_n}$. The total electric potential at the position (x,y=0) is the sum of the potential from each polar, as below:

$$\varphi = \frac{1}{4\pi\varepsilon_0} \sum \frac{q_n}{|\vec{r_n}|}$$

Here $q_n$ is a value with a sign corresponding to the polarity of the charge. Considering the value of each $q_n$ is identical, so

$$\varphi = \frac{1}{4\pi\varepsilon_0} \sum \frac{q_n}{|\vec{r_n}|} \propto \sum \frac{1}{|\vec{r_n}|}$$

So

$$\Delta V_{N1} = \varphi|_{x=0\ nm} - \varphi|_{x=-8\ nm}$$

## S2. A possible experiment for validation of the theory

It has been validated that axonal cues are not necessary for the myelin wrapping of oligodendrocytes, though they are still necessary for myelin compaction [1]. It is highly possible that SCs follow the same principle. So we can design an experiment shown in Figure S2 to validate the contribution of the E-field in myelin development. A mesh of silver micro/nano-wires, 0.2-10 μm in diameter, coated with 1 μm thick parylene as an insulating layer is used as a substitute for the axons with varying calibers. When it is partially immersed in the culture medium, the surface potential can be controlled by the applied voltage, as shown. The oligodendrocytes or SCs can both be cultured with nano-wire in the medium, and the myelination process can be observed by varying the applied voltage. Based on our theory, several phenomena can be predicted as follow:

1. The minimum diameter of the myelinated wire decreases with the increasing amplitude of the negative voltage.

2. When the positive voltage is applied, the myelination process will be inhibited for all wires.

3. If a negative voltage is applied to induce the myelination first, the post-applied positive voltage can induce demyelination (Figure S2(b)).

4. There will be a threshold voltage, $V_{N1\text{-}T'}$, to initiate the myelination process.
5. There will be another threshold voltage, $V_{N2\text{-}T'}$, to initiate the demyelination.

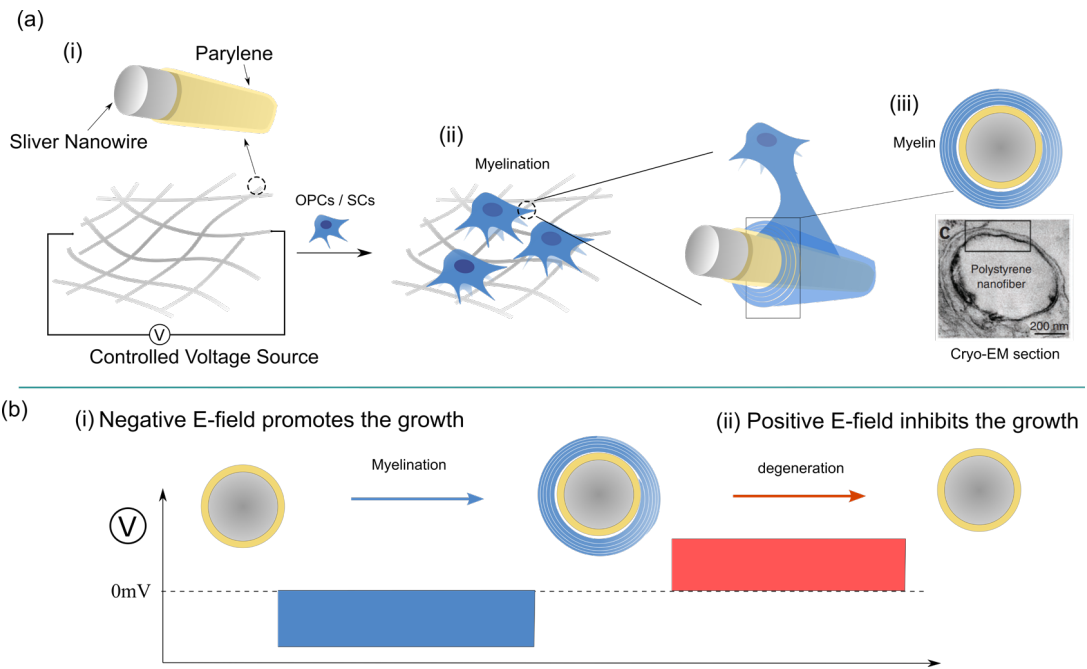

Figure S2. A designed experiment for validation of Hypothesis-E: (a) the experimental setup; (b) Modulate the myelination process by controlling the E-field of the nano-wire.